\documentclass[aps,pre,graphicx,reprint,superscriptaddress]{revtex4-1}

\usepackage{amsmath}
\usepackage{amsfonts}
\usepackage{graphicx}
\usepackage{amssymb}
\usepackage{bm}
\usepackage{braket}
\usepackage{hyperref}
\usepackage{xcolor}

\begin{document}

\title{Solitary zonal structures in subcritical drift waves: a minimum model}

\author{Yao Zhou}
\email[]{yaozhou@princeton.edu}
\affiliation{Princeton Plasma Physics Laboratory, Princeton, New Jersey 08543, USA}

\author{Hongxuan Zhu}
\affiliation{Princeton Plasma Physics Laboratory, Princeton, New Jersey 08543, USA}
\affiliation{Department of Astrophysical Sciences, Princeton University, Princeton, NJ 08544, USA}

\author{I.~Y.~Dodin}
\affiliation{Princeton Plasma Physics Laboratory, Princeton, New Jersey 08543, USA}
\affiliation{Department of Astrophysical Sciences, Princeton University, Princeton, NJ 08544, USA}

\date{\today}

\begin{abstract}
Solitary zonal structures have recently been identified in gyrokinetic simulations of subcritical drift-wave (DW) turbulence with background shear flows. However, the nature of these structures has not been fully understood yet. Here, we show that similar structures can be obtained within a reduced model, which complements the modified Hasegawa--Mima equation with a generic primary instability and a background shear flow. We also find that these structures can be qualitatively reproduced in the modified Hasegawa--Wakatani equation, which subsumes the reduced model as a limit. In particular, we illustrate that in both cases, the solitary zonal structures approximately satisfy the same ``equation of state'', which is a local relation connecting the DW envelope with the zonal-flow velocity. Due to this generality, our reduced model can be considered as a minimum model for solitary zonal structures in subcritical DWs. 

\end{abstract}

\maketitle

\section{Introduction}
Shear flows in magnetically confined plasmas have long been a subject of extensive research due to their ability to regulate turbulence and transport \cite{Terry2000}. \textcolor{black}{In the presence of a background shear flow}, linear modes may transiently grow but ultimately decay, so the plasma is formally stable to small perturbations. Nonetheless, perturbations with sufficiently large amplitudes can develop nonlinearly into what is called \textcolor{black}{subcritical turbulence}. Interestingly, radially propagating coherent structures are often observed in gyrokinetic simulations of \textcolor{black}{subcritical drift-wave (DW) turbulence}. Examples include the so-called avalanches \cite{Candy2003,McMillan2009} and the recently reported solitary zonal structures \cite{VanWyk2016,VanWyk2017,McMillan2018}. [Here, ``solitary'' means propagating at a (roughly) constant speed while maintaining a (roughly) constant shape;  \textcolor{black}{``zonal structures'' contain DWs with radial envelopes and zonal flows (ZFs), but not zonal currents, etc.}] These structures are important in that they can induce transport that is not diffusive but rather ballistic; yet, their nature has not been fully understood. This calls for development of reduced models that can elucidate the underlying basic physics.

A reduced model for solitary zonal structures in subcritical turbulence has been proposed based on a plasma interchange model \cite{Pringle2017}. However, its direct relevance to DW turbulence is unclear, because its modes do not have a real diamagnetic frequency, which is an essential feature of DWs \cite{Horton1999}. 
One might expect that a more relevant model could be based on the modified Hasegawa--Mima equation (mHME) \cite{Chandre2014,Dewar2007}, which is usually considered to be the simplest DW model. 
This is endorsed by the fact that the mHME can indeed support solitary zonal structures, which are approximately nonlinear Schr\"odinger (NLS) solitons \cite{Guo2009,Jovanovic2010,Zhou2019}. \textcolor{black}{Nevertheless, these ``NLS solitons'' deteriorate when a background shear flow is imposed, since the mHME does not have a primary instability to counteract the effect of the shear. Hence, the mHME needs further adaptation in order to describe subcritical DWs and the solitary zonal structures therein.}

In this paper, we propose a reduced model for subcritical DWs by complementing the mHME with a generic primary instability and a background shear flow. Within this model, we readily obtain solitary zonal structures resembling those identified in gyrokinetic simulations \cite{McMillan2018}. While these subcritical solitons have smaller widths and larger amplitudes than the aforementioned NLS solitons, they approximately satisfy the same ``equation of state'', which is a local relation connecting the DW envelope with the ZF velocity. In addition, we find that these results can be qualitatively reproduced in the modified Hasegawa--Wakatani equation (mHWE) \cite{Numata2007}, which subsumes our reduced model as a limit. Therefore, our reduced model can be considered as a minimum model for studying solitary zonal structures in subcritical DWs.

This paper is organized as follows. In Sec.\,\ref{soliton}, we briefly review the NLS solitons in the mHME. In Sec.\,\ref{degradation}, we discuss how these solitons deteriorate in the presence of background shear flows. In Sec.\,\ref{restoration}, we introduce our minimum model and describe the features of the subcritical solitons that it supports. These are the main results of our paper. In Sec.\,\ref{HWS}, we show that these results can be qualitatively reproduced in the mHWE. Our results are summarized and discussed in Sec.\,\ref{discussion}.

\section{NLS solitons in the mHME}\label{soliton}
\subsection{mHME and quasilinear approximation}
First, let us consider DWs within the mHME \cite{Chandre2014,Dewar2007}, which is the simplest yet useful two-dimensional slab model that captures many basic effects of interest. In a dimensionless form, the mHME can be written as
\begin{subequations}\label{mHME}
\begin{gather}
\partial_t w + \mathbf{v}\cdot\nabla w - \beta\partial_y\phi =0,\label{HME}\\
  w \doteq \nabla^2\phi -\tilde{\phi},\label{vorticity}
\end{gather}
\end{subequations}
where the functions $w$ and $\phi$ (and $\tilde{\phi}$ too; see below) are considered on the plane with coordinates $\mathbf{x}\equiv(x,y)$. (The symbol $\doteq$ denotes definitions.) A uniform magnetic field $\mathbf{B}$ is applied perpendicularly to this plane. The gradient of the plasma density $n_0$ is in the radial ($x$) direction, and is parameterized by a (positive) constant $\beta\doteq a/L_{n}$, where $a$ is some system size (e.g., the minor radius of a tokamak) and $L_{n}\doteq (- \mathrm{d}\ln n_0/\mathrm{d}x)^{-1}$ is the local scale length of the density gradient. The ZF velocity is in the poloidal ($y$) direction. Time $t$ is normalized by the transit time $a/c_\text{s}$, where $c_\text{s}$ is the sound speed. Space is normalized by the ion sound radius $\rho_\text{s}\doteq c_\text{s}/\Omega_\text{ci}$, where $\Omega_\text{ci}$ is the ion gyro-frequency. The electrostatic potential $\phi(t,\mathbf{x})$ is normalized by $T_\text{e}\rho_\text{s}/(ea)$, where $e$ is the unit charge and $T_\text{e}$ is the electron temperature.
Accordingly, $\mathbf{v}\doteq(-\partial_y\phi,\partial_x\phi)$ is the $\mathbf{E}\times\mathbf{B}$ velocity.

In the mHME, the definition of the generalized vorticity $w$ \eqref{vorticity} involves separating the total $\phi$ into the zonal component $\langle{\phi}\rangle$ and non-zonal component $\tilde{\phi}$. The former is the ``zonal average'' of $\phi$, $\langle{\phi}\rangle\doteq\int\mathrm{d}y\,\phi/L_y$ (where $L_y$ is the system length in $y$), and corresponds to the ZF. The latter is the fluctuating component, $\tilde{\phi}\doteq\phi-\langle{\phi}\rangle$, and corresponds to DWs. The same notations apply to $w$ and $\mathbf{v}$ as well. The mHME differs from the original Hasegawa--Mima equation \cite{Hasegawa1978} in that the adiabatic electron response in $w$ contains only the non-zonal potential $\tilde{\phi}$ rather than the total potential $\phi$. The need for this modification was first identified in Refs.\,\cite{Dorland1993,Hammett1993}.

\textcolor{black}{In studies of DW--ZF interactions, it is common to invoke the so-called quasilinear approximation \cite{Chen2000,Rogers2000,Champeaux2001,Jenko2006,Gallagher2012,Srinivasan2012,Parker2013,Connaughton2015,St-Onge2017,Dewar2007,Guo2009,Jovanovic2010,Zhou2019,Zhu2019} for simplicity. This approximation amounts to neglecting the direct couplings between DWs such as those described in Ref.\,\cite{Gurcan2010} while keeping their indirect couplings via ZFs. On various occasions, quasilinear modelings of ZF dynamics have been shown to reproduce many of the basic features of nonlinear modelings \cite{Zhou2019,Srinivasan2012,Zhu2019}, which makes them preferable for studying certain aspects of DW--ZF interactions.} The non-zonal and zonal components of the quasilinear mHME are given by, respectively,
\begin{subequations}\label{QL}
\begin{gather}
\partial_t\tilde{w}+U\partial_y\tilde{w}-(\beta+U'')\partial_y\tilde{\phi}=0,\label{QLNZ}\\
\partial_t U -\partial_x\langle{\partial_x\tilde{\phi}  \partial_y\tilde{\phi}}\rangle=0.\label{QLZ}
\end{gather}
\end{subequations}
For convenience, we introduce the ZF velocity $U(t,x)\doteq \langle v_y\rangle =  \partial_x\langle\phi\rangle$ here, with $U''\doteq\partial^2_xU$.  

The poloidal wavenumber $k_y$ of a DW $\tilde{w}=\text{Re}[\varpi(t,x)e^{i{k_y}y}]$ is a constant of motion in the quasilinear mHME \eqref{QL}. Hence, we can further restrict our scope to such DWs that are monochromatic in $y$. By denoting $\tilde{\phi}=\text{Re}[\varphi(t,x)e^{i{k_y}y}]$ such that $\varpi = (\partial_x^2-k_y^2-1)\varphi$, we obtain the one-dimensional (1D) quasilinear mHME:  
\begin{subequations}\label{QL1}
\begin{gather}
\partial_t\varpi+ik_yU\varpi-ik_y(\beta+U''){\varphi}=0,\label{QL1NZ}\\
\partial_t U -k_y\textrm{Im}({\varphi}^*{\partial_x^2{\varphi}  })/2=0.\label{QL1Z}
\end{gather}
\end{subequations}
Here, ${\varphi}^*$ is the complex conjugate of ${\varphi}$, and the factor $1/2$ is due to zonal averaging. Equation \eqref{QL1} represents the basic model that our study builds upon.

\subsection{NLS equation and solitons} 
Equations \eqref{mHME}, \eqref{QL}, and \eqref{QL1} allow monochromatic DWs $\varpi=\psi_0e^{i{k_x}{x}-i\Omega t}$ as exact nonlinear solutions, where $\psi_0$ is a constant, $k_x$ is the radial wavenumber, and $\Omega\doteq\beta k_y/\bar{k}^2$ is the DW frequency, with $\bar{k}^2\doteq 1+k_x^2+k_y^2$. 
Now, let us consider a quasi-monochromatic DW with slow radial modulation only, i.e., $\varpi=\psi(t,x)e^{i{k_x}{x}-i\Omega t}$ with $|\partial_x\ln\psi|\ll |k_x|$. Assuming also that $|\psi|$ is small, one can show that $U$ and $|\psi|$ are connected by a simple ``equation of state" \cite{Zhou2019}
\begin{align}
U \approx|\psi|^2/(4\beta)=\langle\tilde{w}^2\rangle/(2\beta), \label{EOS}
\end{align}
and furthermore, the governing equation for $\psi$ is \cite{Champeaux2001,Dewar2007}
\begin{align}
i(\partial_t +v_\text{g} \partial_x)\psi \approx -(\chi/2)\partial_x^2\psi+k_y|\psi|^2 \psi/(4\beta).\label{NLSE}
\end{align}
Here, $v_\text{g}\doteq\partial\Omega/\partial k_x$ is the radial group velocity and $\chi\doteq\partial^2\Omega/\partial k_x^2$. More explicitly, $v_\text{g}= -2\beta k_xk_y/\bar{k}^4$ and $\chi= (2\beta k_y/\bar{k}^6)(4k_x^2-\bar{k}^2)$. Detailed derivations of Eqs.\,\eqref{EOS} and \eqref{NLSE} can be found in Ref.\,\cite{Zhou2019} and the references therein.

Since Eq.\,\eqref{NLSE} has the form of a NLS equation, a DW packet can be considered as an effective quantum particle (``drifton''), for which $\psi$ serves as a state function. Also, as a NLS equation, Eq.\,\eqref{NLSE} has the usual soliton solution
\begin{align}
\psi(t,x)=2\eta\sqrt{-\frac{\beta\chi}{k_y}}\frac{\exp{(i\chi\eta^2t/2)}}{\cosh[\eta(x-v_\text{g}t)]}.\label{psiS}
\end{align}
Here, the soliton inverse width $\eta$ is a free parameter that also determines the soliton amplitude. The corresponding approximate solution to the mHME is then given by $\tilde{w}=\text{Re}(\psi e^{i{k_x}{x}+ik_y y-i\Omega t})$, together with Eq.\,\eqref{EOS}. Below, we use the term ``NLS solitons'' to denote such solitary solutions to the mHME specifically, rather than those to the NLS equation \eqref{NLSE} in general.
In Ref.\,\cite{Zhou2019}, it is shown numerically that NLS solitons can be generated via the modulational instability of (quasi-) monochromatic DWs.

\begin{figure}
  \centering
 \includegraphics[width=\columnwidth]{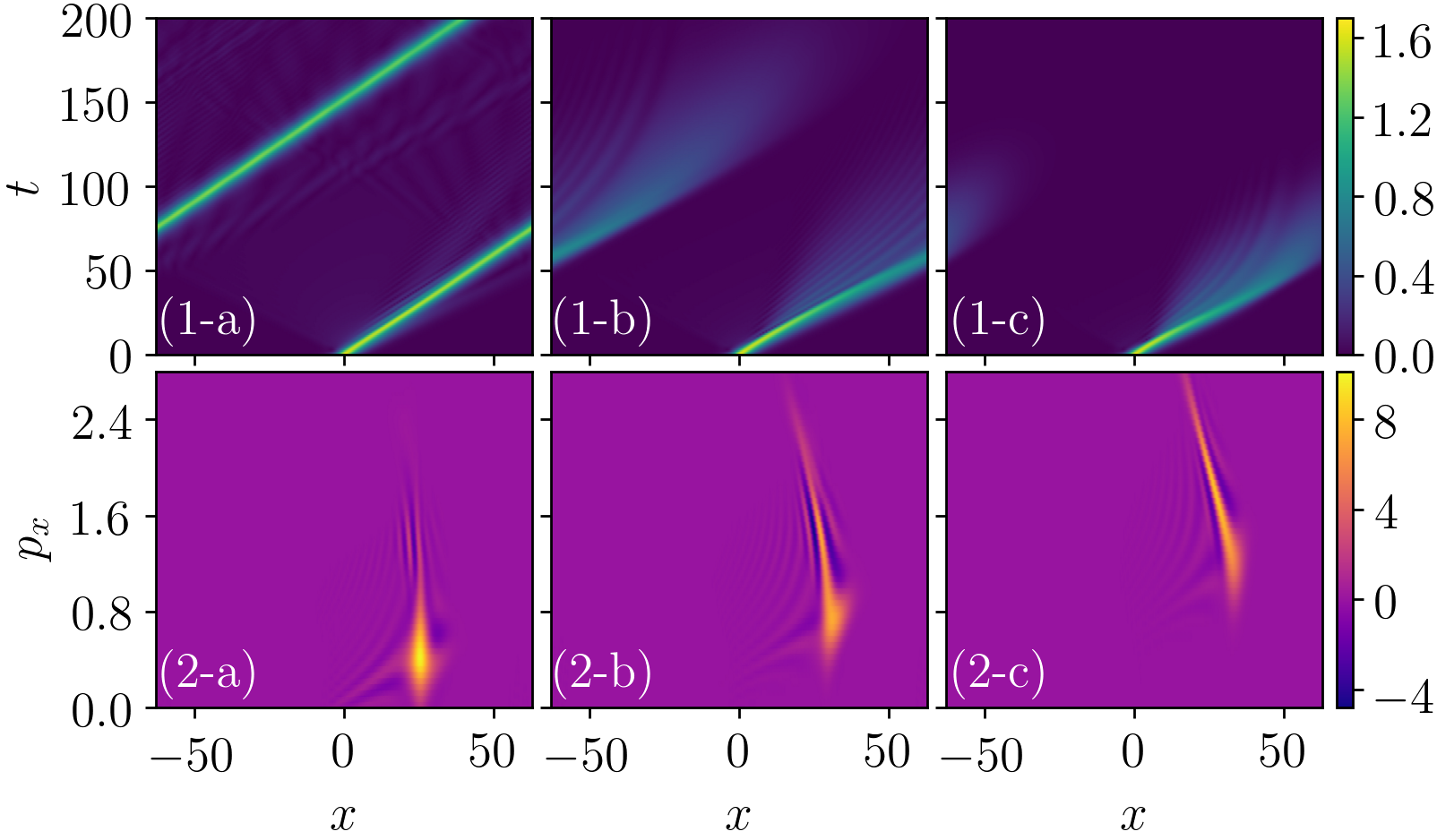}
 \caption{Quasilinear mHME simulations initialized with a NLS soliton \eqref{psiS} ($\beta=5$, $k_x=0.5$, $k_y=-1$, $\eta=0.5$). The columns correspond to various values of the background flow shear: (a) $S= 0$, (b) $S= 0.02$, and (c) $S= 0.04$. Row~1 shows the spatial-temporal evolution of the DW envelope $\sqrt{\langle\tilde{w}^2\rangle}$. Row 2 shows the Wigner function $W(x,p_x)$ at $t = 30$.}\label{mHMEhistory}
\end{figure}

Although $\eta\ll |k_x|$ is assumed in the derivation of the NLS soliton \eqref{psiS}, in practice, structures of this form remain solitary even when $\eta\sim |k_x|$. This can be seen in the quasilinear mHME simulation shown in Fig.\,\ref{mHMEhistory}(1-a), which is initialized with a NLS soliton with $k_x=\eta=0.5$. [A snapshot of the envelope of this soliton can be found in Fig.\,\ref{ZF}, which verifies the equation of state \eqref{EOS}.] Still, when $\eta$ is sufficiently larger than $k_x$, the solitary behavior of the zonal structure eventually breaks down \cite{Zhou2019}.

Within the quantum analogy, NLS solitons can be regarded as quasi-monochromatic drifton condensates. A particularly transparent way to illustrate this is by using the Wigner function \cite{Wigner1932}
\begin{align}
W(t,x,{p_x})\doteq\int\mathrm{d}s\,e^{-i{p_x}{s}}\varpi\left(t,{x}+\frac{{s}}{2}\right)\varpi^*\left(t,{x}-\frac{{s}}{2}\right),\label{WF}
\end{align}
where $p_x$ is the coordinate in the DW radial momentum (wavenumber) space. The Wigner function $W$ can be considered as a quasi-probability distribution of driftons (DW quanta) in phase space. (The prefix ``quasi'' denotes the fact that as a quantumlike particle, a drifton has well-defined phase-space coordinates only in the geometrical-optics limit, while the definition of $W$ extends also beyond this limit.) In Fig.\,\ref{mHMEhistory}(2-a), a snapshot of the Wigner function is shown, and it can be seen that the DW quanta are localized in both the coordinate space and the momentum space (specifically, at $p_x\sim k_x$). In this paper, we use the Wigner function only as a visualization tool. However, it can also be directly used to model the (statistical) dynamics of DWs within the so-called Wigner--Moyal formulation \cite{Ruiz2016}, as done in Refs.\,\cite{Zhu2018,Zhou2019,Zhu2019}. \textcolor{black}{Notably, this formulation captures essential ``full-wave'' effects such as diffraction, which are neglected in wave-kinetic approaches based on the ray approximation \cite{Smolyakov1999,Kaw2002,Trines2005,Parker2016,Zhu2018b,Ruiz2019}. Hence, the former can model NLS solitons whereas the latter cannot \cite{Zhou2019,Diamond2005}.}

\section{Degradation of DW packets due to shear flow}\label{degradation}
In this section, we illustrate the effect of background shear flows on NLS solitons. \textcolor{black}{Following Refs.\,\cite{McMillan2018,Pringle2017}, let us introduce a background flow $U_0=Sx$ with some constant shear $S$.} The non-zonal component \eqref{QL1NZ} of the 1D quasilinear mHME then becomes
\begin{gather}
\partial_t\varpi+ik_y(U+Sx)\varpi-ik_y(\beta+U''){\varphi}=0,\label{QLNZS}
\end{gather}
\textcolor{black}{while the zonal component \eqref{QL1Z} still determines the ZF velocity $U$ self-consistently}. It is easy to find that the exact monochromatic DW solution now takes the form
 \begin{gather}
\varpi=\psi_0\exp\left[{i{k_x}(t){x}-i\int\Omega(t)\,\mathrm{d}t}\right],
\end{gather}
where
 \begin{subequations}\label{kOmega}
\begin{gather}
k_x(t)=K_x-Stk_{y}, \\
\Omega(t)=\beta k_y/[1+{k_x}(t)^2+k_y^2],
\end{gather}
\end{subequations}
and $K_x$ is the initial radial wavenumber. The fact that the radial wavenumber $k_{x}(t)$ changes linearly with time while the poloidal wavenumber $k_y$ stays constant is a typical effect of the background shear flow. (It exactly applies to many other systems such as the nonlinear mHME \eqref{mHME}; see Appendix\,\ref{shearingbox}.)

It turns out that this effect on monochromatic DWs also qualitatively extends to weakly nonlinear quasi-monochromatic DW packets, including NLS solitons. In Fig.\,\ref{mHMEhistory}, we show results from quasilinear mHME simulations initialized with a NLS soliton \eqref{psiS} for different values of the flow shear $S$. (The implementation of these simulations is described in Appendix\,\ref{shearingbox}.)
The snapshots of the Wigner function in row 2, which are taken at the same $t$ for various $S$, demonstrate that the background shear flow does change the characteristic radial wavenumber $k_x$ of the DW packet, and the change increases with $S$. As $t$ increases, the group velocity $v_g=\partial\Omega/\partial k_x$ decreases, and hence the propagation slows down. While the background shear flow itself does not dissipate the DW quanta \textcolor{black}{(i.e., does not affect the conservation of the DW enstrophy $Z_\mathrm{DW}\doteq \int\mathrm{d}x\,\langle\tilde{w}^2\rangle/2$)}, it keeps increasing $|k_x|$ such that the dissipation excluded in Eq.\,\eqref{QL} eventually becomes non-negligible. (A small amount of hyper-viscosity is included in the simulations shown in Fig.\,\ref{mHMEhistory}.) Meanwhile, there is no source in the mHME to replenish the DW quanta, so the DW packet inevitably deteriorates [Fig.\,\ref{mHMEhistory}(b-c)]. That is, NLS solitons cannot survive when a background shear flow is imposed in the mHME.

\section{Subcritical solitons sustained by primary instability}\label{restoration}
Following Ref.\,\cite{Guo2009}, we introduce a source (and a sink) to the 1D quasilinear mHME \eqref{QL1} by complementing the latter with a generic primary instability (including explicit dissipation). Also keeping the background shear flow in Eq.\,\eqref{QLNZS}, we obtain a reduced model for subcritical DWs:
\begin{subequations}\label{QLP}
\begin{gather}
\partial_t\varpi+ik_y(U+Sx)\varpi-ik_y(\beta+U''){\varphi}=\hat\gamma\varpi,\label{QLNZP}\\
\partial_t U -k_y\textrm{Im}({\varphi}^*{\partial_x^2{\varphi}  })/2=\hat\gamma U.\label{QLZP}
\end{gather}
\end{subequations}
Here, the operator $\hat\gamma$ is given by $\hat\gamma=\gamma(\hat{k}_x,k_y)$, with $\hat{k}_x\doteq-i\partial_x$ being the radial momentum (wavenumber) operator and $\gamma({k}_x,k_y)$ the linear growth rate of a monochromatic DW. \textcolor{black}{Unlike the background shear flow, this source term changes the DW enstrophy (quanta) but not the wavenumber}. In general, $\gamma$ should be positive at small $|k_x|$ and negative at large $|k_x|$, such that a linear perturbation may grow transiently but ultimately decays as $|k_x|$ increases. The system is therefore linearly stable, but perturbations with sufficient amplitudes can still develop nonlinearly into turbulence. 

In practice, we adopt the following simple \textit{ad hoc} model of $\gamma$ (in this section only): 
\begin{gather}\label{gammaPI}
\gamma({{k}_x,k_y})=\sigma| k_y| - Dk^2,
\end{gather}
where $\sigma$ and $D$ are positive constants and $k^2\doteq k_x^2+k_y^2$. We do not expect the specific form of $\gamma$ to qualitatively impact our results. In Sec.\,\ref{HWS}, we show that the mHWE, which has a growth rate different from Eq.\,\eqref{gammaPI}, produces qualitatively similar results to those here.

\begin{figure}
  \centering
 \includegraphics[width=\columnwidth]{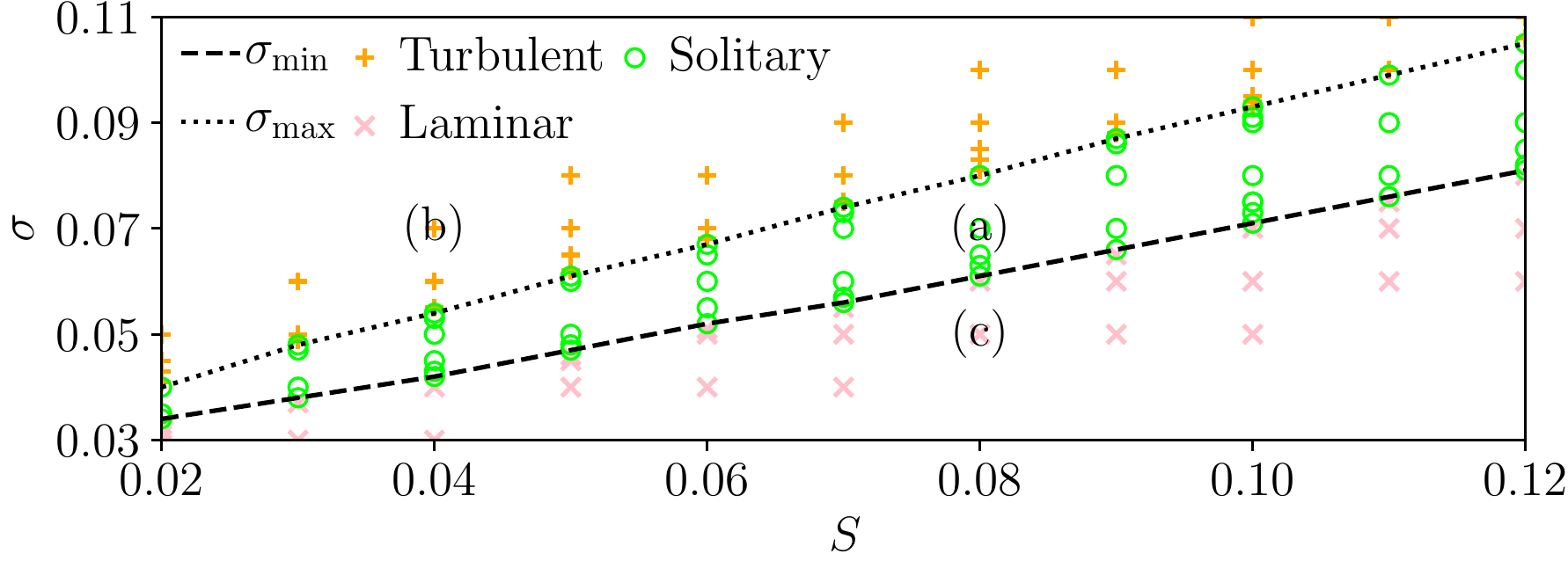}
 \caption{Results from simulations of the reduced model \eqref{QLP} ($\beta=5$, $D=0.02$) with various $\sigma$ and $S$. The initial condition is given by Eq.\,\eqref{psiS} ($k_x=0$, $k_y=-1$, $\eta=0.5$). The maximum (dotted) and minimum (dashed) values of $\sigma$ for the solutions to be solitary are plotted vs.~$S$. The points (a)-(c) correspond to the three columns in Fig.\,\ref{Edge}, respectively.}\label{sigmaS}
\end{figure}

We perform series of simulations of the reduced model \eqref{QLP} with various $\sigma$ and $S$, while fixing $D$ and the initial condition for $\varpi$ ($\psi$). The results of the simulations are summarized in Fig.\,\ref{sigmaS}. Note that if $S=0$, the initial condition we use would correspond to a stationary zonal structure with $k_x=0$. However, the background shear flow changes $k_x$ such that the zonal structure starts propagating. For a given $S$, when the primary instability is moderate ($\sigma_{\text{min}}\le\sigma\le\sigma_{\text{max}}$), the zonal structure becomes and remains solitary for the duration of the simulations (up to $t=1000$). When the primary instability is too strong ($\sigma>\sigma_{\text{max}}$), the zonal structure keeps growing and starts avalanching, and eventually the system becomes turbulent. When the primary instability is too weak ($\sigma<\sigma_{\text{min}}$), the zonal structure cannot be sustained and ultimately deteriorates, similarly to the cases in Fig.\,\ref{mHMEhistory}(b-c). Both $\sigma_{\text{min}}$ and $\sigma_{\text{max}}$ scale roughly linearly with $S$, as shown in Fig.\,\ref{sigmaS}. 

\begin{figure}
  \centering
 \includegraphics[width=\columnwidth]{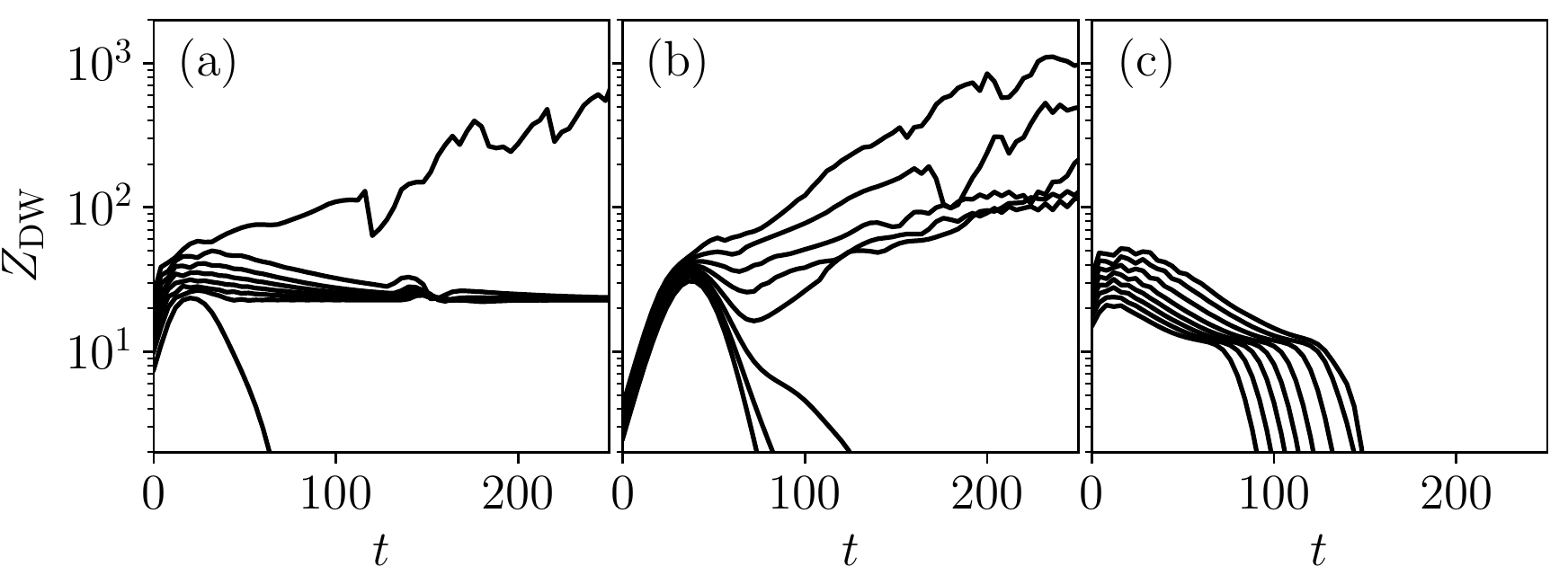}
 \caption{The time history of the DW enstrophy $Z_\mathrm{DW}$ for different pairs of $(\sigma, S)$: (a) $(0.07, 0.08)$, (b) $(0.07, 0.04)$, and (c) $(0.05, 0.08)$. The simulations have the same setup as those in Fig.\,\ref{sigmaS}, except that the amplitude of the initial condition $\eta$ is varied in each column. Consequently, not all structures in (a) are solitary and not all structures in (b) are unstable.}\label{Edge}
\end{figure}

Initial conditions, their amplitudes in particular, typically play an important role in subcritical systems. Nevertheless, the qualitative features of Fig.\,\ref{sigmaS} depends only weakly on the specific initial condition used therein. This is shown in Fig.\,\ref{Edge}, where we choose three representative points from Fig.\,\ref{sigmaS} and then vary the amplitude of the initial condition $\eta$. In Fig.\,\ref{Edge}(a), for a sizable range of $\eta$, the system settles into (almost) the same final stage with a solitary DW; only when $\eta$ is sufficiently large (small) does the system become turbulent (laminar). In Fig.\,\ref{Edge}(b), the system tends to become turbulent unless $\eta$ is really small, in which case the system turns laminar. Here, no robust solitary solution could be found, except maybe transient ones near the fine edge between the turbulent and laminar states, similar to the ``edge of chaos" scenarios discussed in Refs.\,\cite{Pringle2017,McMillan2018}. In Fig.\,\ref{Edge}(c), even for initial conditions with reasonably large $\eta$, the DW packets still decay, and the system ends up laminar.

\begin{figure}
  \centering
 \includegraphics[width=\columnwidth]{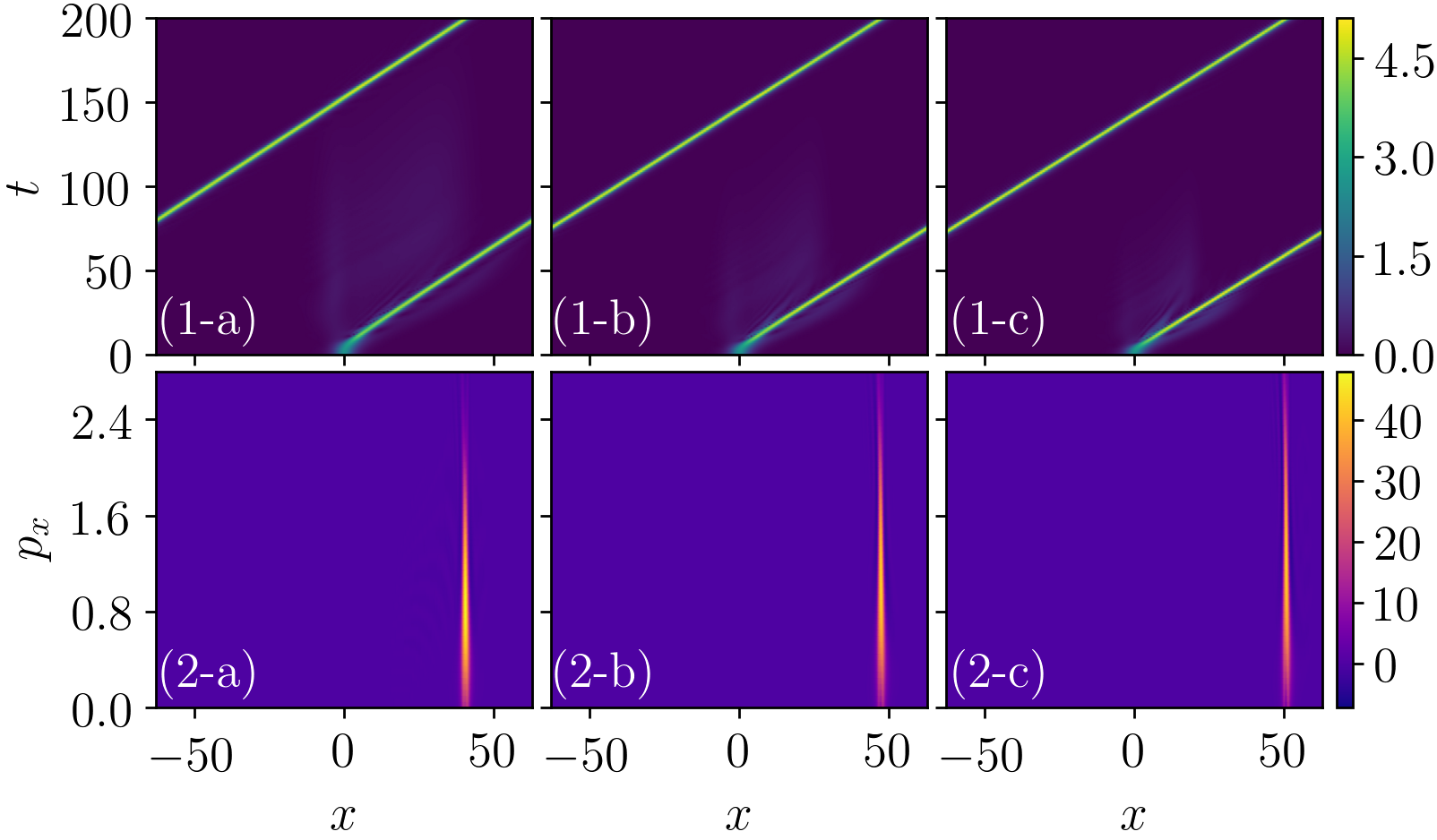}
 \caption{Examples of solitary zonal structures obtained from selected simulations in Fig.\,\ref{sigmaS}. The columns correspond to different pairs of $(\sigma, S)$: (a) $(0.05, 0.04)$, (b) $(0.06, 0.06)$, and (c) $(0.07, 0.08)$. Row 1 shows the spatial-temporal evolution of the DW envelope $\sqrt{\langle\tilde{w}^2\rangle}$. Row 2 shows the Wigner function $W(x,p_x)$ at $t = 200$.}\label{shearhistory}
\end{figure}

In Fig.\,\ref{shearhistory}, we present examples of solitary zonal structures obtained with different pairs of $(\sigma, S)$. By comparing the spatial-temporal evolution of the DW envelope and the snapshots of the Wigner function, we can see that these structures are quite similar among themselves. For clarity, we refer to these structures as ``subcritical solitons'' in this paper, \textcolor{black}{since they are visibly different from the NLS soliton in Fig.\,\ref{mHMEhistory}(a). That is, subcritical solitons have smaller widths, and in turn, larger amplitudes than NLS solitons. Accordingly, the Wigner function of a subcritical soliton is more localized in the coordinate space while more spread out in the momentum space. Note that 
without primary instabilities and background shear flows, the mHME cannot support NLS solitons with such large inverse widths, because it is not approximated well by the NLS equation \eqref{NLSE} in this case \cite{Zhou2019}}.


\begin{figure}
  \centering
 \includegraphics[width=\columnwidth]{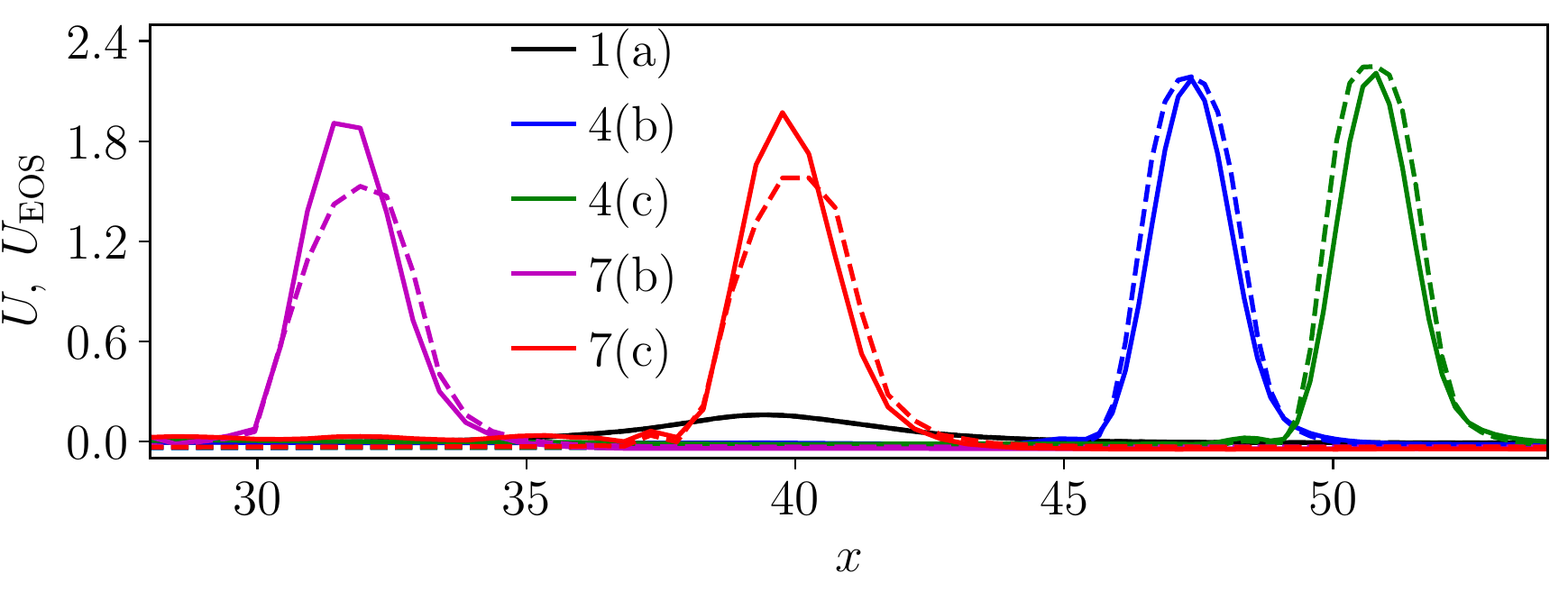}
 \caption{Snapshots of the actual ZF velocity $U$ (solid) and the ZF velocity calculated from the equation of state \eqref{EOS} [$U_\mathrm{EOS}\doteq \langle\tilde{w}^2\rangle/(2\beta)$, dashed] at the end of various simulations, specifically, a NLS soliton [Fig.\,\ref{mHMEhistory}(a)], and the subcritical solitons in the reduced model [Fig.\,\ref{shearhistory}(b)-(c)] and the mHWE [Fig.\,\ref{mHWEhistory}(b)-(c)].}\label{ZF}
\end{figure}

Meanwhile, subcritical solitons and NLS solitons still share the similarity in that they are both nonlinearly sustained by DW--ZF interactions, in which the ZF acts as a self-trapping potential. Also, perhaps surprisingly and remarkably, subcritical solitons satisfy the equation of state \eqref{EOS} of NLS solitons, as found numerically. This can be seen in Fig.\,\ref{ZF}, where snapshots of the ZF velocity are shown to agree well with those calculated from Eq.\,\eqref{EOS}, even though subcritical solitons have smaller widths and larger amplitudes. In this sense, both NLS solitons and subcritical solitons can be categorized as DW--ZF solitons. 

\textcolor{black}{In summary, subcritical solitons are sustained by an intricate balance of the following effects combined: the generation and dissipation of DW quanta at small and large $|k_x|$ by the source and the sink, respectively; the channeling of DW quanta from small to large $|k_x|$ by the background shear flow; the self-trapping via DW--ZF interactions; and the diffraction of the DW envelope. Subcritical solitons are not possible if any of these effects is absent. The reduced model \eqref{QLP} contains all these effects and hence can be considered as a minimum model for solitary zonal structures in subcritical DWs. }

\section{Subcritical solitons in the mHWE}\label{HWS}
In order to show the physical relevance of the reduced model \eqref{QLP}, let us examine whether its results can be reproduced with more complex models that subsume it as a limit. One such model is the mHWE \cite{Numata2007}, in which a modification similar to the one in the mHME (see Sec.\,\ref{soliton}) is applied to the original Hasegawa--Wakatani equation \cite{Hasegawa1983}. The mHWE reads
\begin{subequations}\label{mHWE}
\begin{gather}
\partial_t w+\mathbf{v}\cdot\nabla{w}-\beta\partial_y\phi=D\nabla^2w,\\
\partial_t n+\mathbf{v}\cdot\nabla{n}+\beta\partial_y\phi=\alpha(\tilde{\phi}-\tilde{n})+D\nabla^2 n.
\end{gather}
\end{subequations}
Here, $n$ is the perturbed density normalized by $n_0\rho_s /a$, $w \doteq \nabla^2\phi-{n}$ is the generalized vorticity, $\alpha$ is the adiabaticity parameter, and the form of dissipation is chosen to be the same as that in Sec.\,\ref{restoration}. In the so-called adiabatic limit, where $D\rightarrow 0 $ and $\alpha\rightarrow\infty$, we should have $\tilde{n}\rightarrow\tilde{\phi}$. Accordingly, one can deduce that $\partial_t\langle n\rangle\rightarrow 0$ and hence adopt $\langle n\rangle\rightarrow 0$. Then, with $w \rightarrow \nabla^2\phi-\tilde{\phi}$, the mHWE \eqref{mHWE} formally converges to the mHME \eqref{mHME}. Notably, while the results presented in this section corroborate this correspondence between the mHWE and the mHME, the robustness of such convergence in general is a subtle issue and may not be guaranteed \cite{Majda2018}.

\begin{figure}
  \centering
 \includegraphics[width=\columnwidth]{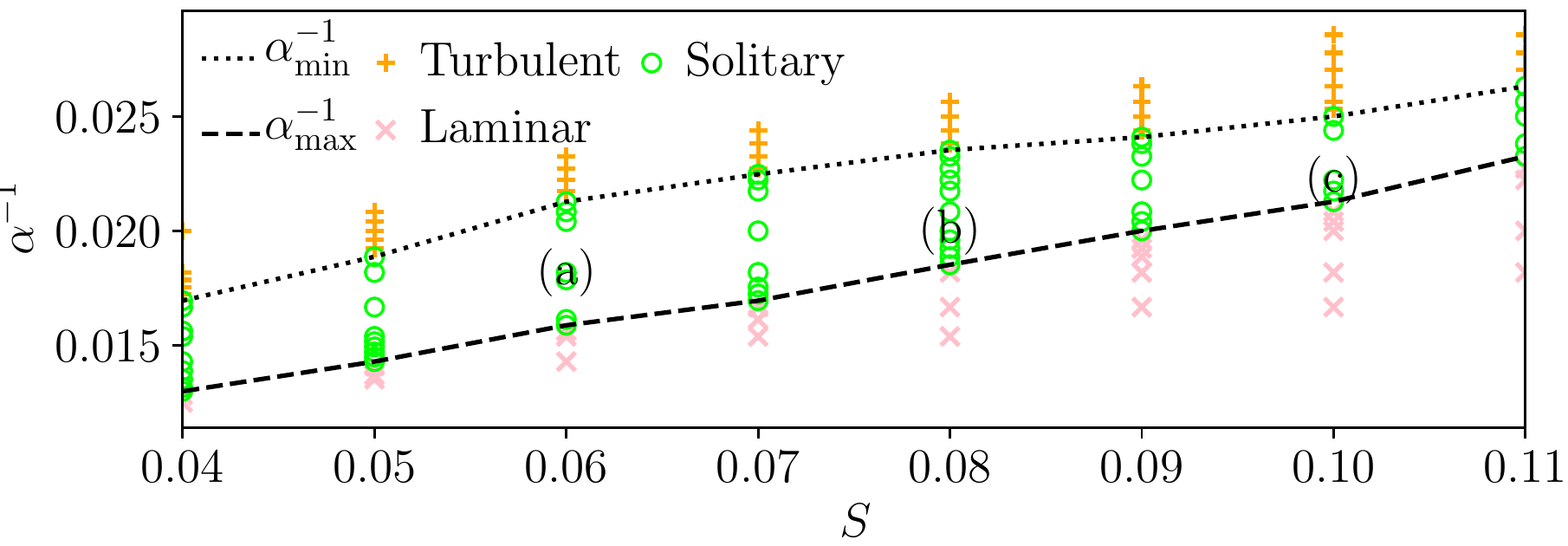}
 \caption{\textcolor{black}{Results from mHWE simulations ($\beta=5$, $D=0.02$) with background shear flows and various $\alpha$ and $S$. The initial condition is given by Eq.\,\eqref{psiS} ($k_x=0$, $k_y=-1$, $\eta=0.6$, $n=\tilde\phi$). The inverse of the maximum (dashed) and minimum (dotted) values of $\alpha$ for the solutions to be solitary are plotted vs.~$S$. The points (a)-(c) correspond to the three columns in Fig.\,\ref{mHWEhistory}, respectively.} }\label{alphaS}
\end{figure}

\begin{figure}
  \centering
 \includegraphics[width=\columnwidth]{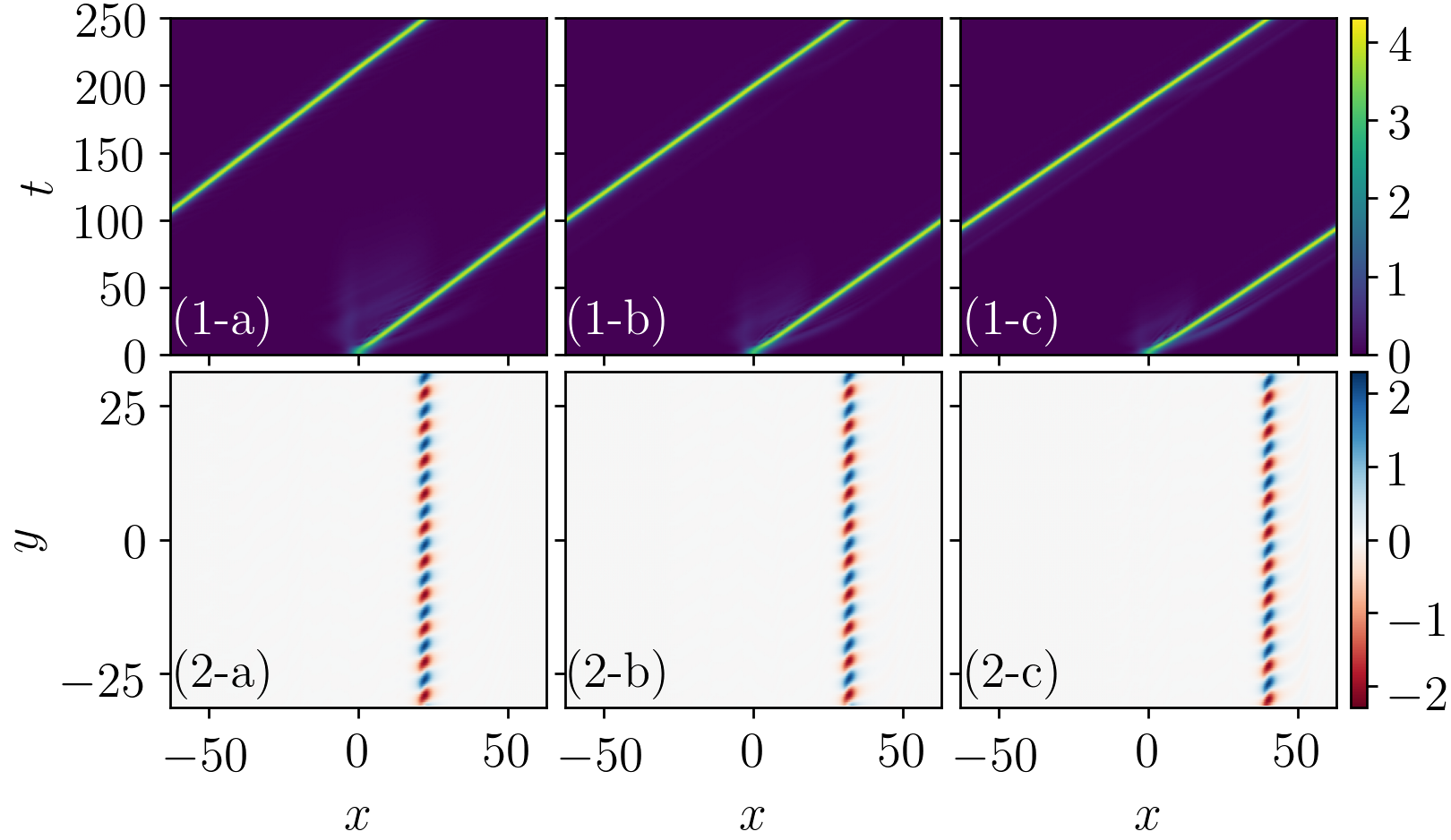}
 \caption{Examples of solitary zonal structures obtained from selected simulations in Fig.\,\ref{alphaS}. The columns correspond to different pairs of $(\alpha, S)$: (a) $(55, 0.06)$, (b) $(50, 0.08)$, and (c) $(45, 0.1)$. Row 1 shows the spatial-temporal evolution of the DW envelope $\sqrt{\langle\tilde{w}^2\rangle}$. Row 2 shows the non-zonal potential $\tilde\phi(x,y)$ at $t = 250$.}\label{mHWEhistory}
\end{figure}

Close to the adiabatic limit, one branch of the dispersion relation of the mHWE reads $\Omega\approx\beta k_y/\bar{k}^2+i\gamma$, with its real part converging to the mHME case. The imaginary part reads
\begin{gather}\label{gammaHW}
\gamma({{k_x},k_y})= {\beta^2k_y^2k^2}/({\alpha\bar{k}^6}) - Dk^2,
\end{gather}
which differs from the simple model \eqref{gammaPI} used in Sec.\,\ref{restoration}. However, as we impose background shear flows in simulations of the mHWE \eqref{mHWE} (the implementation is described in Appendix \ref{shearingbox}), \textcolor{black}{we can still obtain qualitatively similar results. In Fig.\,\ref{alphaS}, we summarize the results of series of simulations with various $\alpha$ and $S$, where $\alpha^{-1}$ plays the same role as $\sigma$ in Fig.\,\ref{sigmaS}, i.e., indicating the strength of the primary instability. For a given $S$, solitary zonal structures can be obtained only when the primary instability is moderate. When the primary instability is too strong or too weak, the system becomes turbulent or laminar, respectively. Examples of the solitary zonal structures obtained are shown in Fig.\,\ref{mHWEhistory}, which resemble those obtained with the reduced model (Fig.\,\ref{shearhistory}). Due to this similarity, we refer to these structures also as subcritical solitons.}


\textcolor{black}{We emphasize that these mHWE simulations are nonlinear (with self-consistent spectra in $k_y$) rather than quasilinear as in the reduced model (with a single $k_y$). Nonetheless, the snapshots in Fig.\,\ref{mHWEhistory} (row 2) clearly show that the structures remain quasi-monochromatic in $y$, which in turn justifies the quasilinear approximation.} In fact, the same feature is also displayed by some solitary zonal structures identified in gyrokinetic simulations \cite{McMillan2018} [Fig.\,3(b) therein]. Deviations of the mHWE from its quasilinear approximation does affect the applicability of the equation of state \eqref{EOS} slightly. In Fig.\,\ref{ZF}, the agreement between the ZF velocity and that calculated from Eq.\,\eqref{EOS} is not as good in the mHWE simulations as those with the reduced model. [In quasilinear mHWE simulations (not shown), by contrast, the agreement \textit{is} as good.] Still, the equation of state roughly captures the local relation between the DW envelope and the ZF velocity, qualifying these structures also as DW--ZF solitons. \textcolor{black}{The fact that the results obtained with the reduced model \eqref{QLP} can be qualitatively reproduced with the more complex mHWE supports that the former can be considered as a minimum model for solitary zonal structures in subcritical DWs.}

\begin{figure}
  \centering
 \includegraphics[width=\columnwidth]{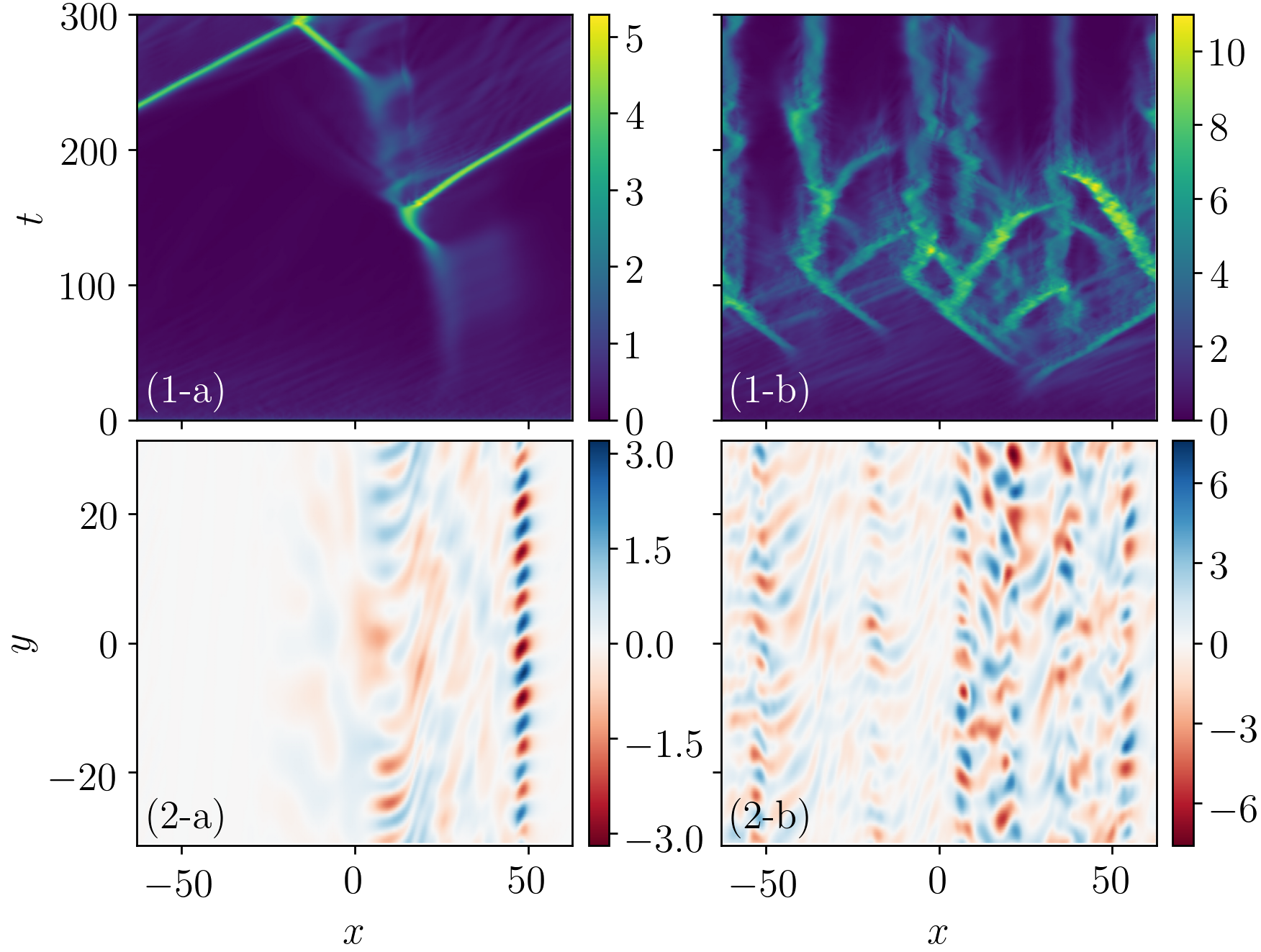}
 \caption{\textcolor{black}{Two mHWE simulations with background shear flows ($\beta=5$, $D=0.02$, $S=0.08$) initialized with identical random perturbations of sufficiently large amplitude: (a) $\alpha=50$, and (b) $\alpha=25$. Row 1 shows the spatial-temporal evolution of the DW envelope $\sqrt{\langle\tilde{w}^2\rangle}$. Row 2 shows the non-zonal potential $\tilde\phi(x,y)$ at (a) $t = 210$ and (b) $t = 70$.}
 }\label{MIhistory}
\end{figure}

One might wonder whether subcritical solitons can naturally emerge from random perturbations rather than the carefully chosen initial conditions that we used so far. The answer appears to be affirmative. In Fig.\,\ref{MIhistory}(a), we show an example of spontaneous subcritical solitons obtained in mHWE simulations. Here, the initial perturbation needs to be of sufficiently large amplitude, because the system is subcritical. This is different from Ref.\,\cite{Zhou2019}, where small perturbations are applied on primary DWs to form NLS solitons in mHME simulations. 
While the snapshot of the solitary structure in Fig.\,\ref{MIhistory}(2-a) exhibits some weak poloidal modulation as well, its clear resemblances to those in Fig.\,\ref{mHWEhistory} (row 2), such as having a dominant poloidal wavenumber, suggest that they are all essentially the same structures.

\textcolor{black}{For comparison, in Fig.\,\ref{MIhistory}(b), we present results from a simulation with a smaller $\alpha$ than in Fig.\,\ref{MIhistory}(a). Since the primary instability is stronger in this case, the propagating structures that are formed do not saturate into solitons, but keep growing and eventually develop into turbulence. Still, features similar to those of subcritical solitons (e.g., nonzero radial wavenumbers) can be seen in Fig.\,\ref{MIhistory}(2-b). These transient structures might be related to the avalanche-like bursts observed in gyrokinetic simulations with background shear flows \cite{Candy2003,McMillan2009}.
}

\section{Summary and discussion}\label{discussion}
In this paper, we propose a minimum model for studying solitary zonal structures in subcritical DWs. This model complements the mHME with a generic primary instability and a background shear flow. The subcritical solitons supported by our minimum model have smaller widths and larger amplitudes than NLS solitons, which are known solutions to the mHME. Nevertheless, we find that these subcritical solitons satisfy the same ``equation of state'' as NLS solitons, which is a local relation that connects the DW envelope with the ZF velocity. Moreover, we show that these results can be qualitatively reproduced in the more complex mHWE, which subsumes our minimum model as a limit.

It could be of interest to pursue future research in the following directions. First, since subcritical solitons seem to be coherent DW packets with dominant wavenumbers, it may be possible that an analytical solution can be derived, similar to the NLS soliton \eqref{psiS}. \textcolor{black}{However, the slow-envelope assumption, which enables the NLS reduction of the mHME, is far from applicable to subcritical solitons. It is likely that a different analytical approach is required.} Second, more careful and systematic comparisons between the subcritical solitons in the minimum model and those found in gyrokinetic simulations could be useful. \textcolor{black}{Third, one could investigate the connection between subcritical solitons and the radially propagating structures induced by self-consistent (rather than background) shear flows, such as those reported recently in Ref.\,\cite{Zhu2020}.}

\acknowledgments
This research was supported by the U.S.~Department of Energy under Contract No.~DE-AC02-09CH11466. The data used for the figures in this article are available at: http://arks.princeton.edu/ark:/88435/dsp015425kd34n.

\appendix
\section{Implementation of shearing-box simulations}\label{shearingbox}
In this appendix, we describe the implementation of numerical simulations with background shear flows, using the nonlinear mHME \eqref{mHME} as an example. The cases with the quasilinear mHME \eqref{QL} and the mHWE \eqref{mHWE} are similar and hence can be straightforwardly inferred.

When a background flow $U_0=Sx$ with constant shear $S$ is imposed, the mHME \eqref{mHME} becomes
\begin{subequations}\label{mHMES}
\begin{gather}
\partial_t{w}+(Sx+\partial_x\phi)\partial_y{w}-(\beta+\partial_x{w})\partial_y{\phi}=0,\\
w=\nabla^2\phi-\tilde{\phi}.
\end{gather}
\end{subequations}
Let us change the coordinates from $(x,y)$ to $(x,Y)$, where $Y\doteq y-Stx$ is a Lagrangian coordinate, and denote $w_0(x,Y,t)=w(x,y,t)$ and $\phi_0(x,Y,t)=\phi(x,y,t)$. The mHME \eqref{mHMES} then becomes
\begin{subequations}\label{mHMESL}
\begin{gather}
\partial_t{w_0}+\partial_x\phi_0\partial_{Y}{w_0}-(\beta+\partial_{x}{w_0})\partial_{Y}{\phi_0}=0,\\
w_0=[(\partial_x-St\partial_{Y})^2+\partial_{Y}^2]\phi_0-\tilde{\phi}_0.
\end{gather}
\end{subequations}
In the so-called ``shearing-box" geometry, the simulation domain is a doubly periodic box in $(x,Y)$, with $(L_x,L_y)$ being the box size. \textcolor{black}{This is a standard local treatment of systems with differential rotations, such as accretion disks \cite{Hawley1995,Squire2015}.} Then, Eq.\,\eqref{mHMESL} can naturally be solved using Fourier-based pseudo-spectral methods (for which we use the \textsc{Dedalus} code \cite{Dedalus}). As discussed in Sec.\,\ref{degradation}, the radial wavenumber of a monochromatic wave will change linearly in time, i.e., $k_x=K_x-Stk_{y}$. In fact, the spectral coordinates of Eq.\,\eqref{mHMESL} exactly correspond to $(K_x,k_y)$. In this sense, this approach essentially uses a spectral grid that constantly shifts in $k_x$. 

A drawback of this Lagrangian approach is that as $t$ increases, the physical grid becomes heavily distorted in $(x,y)$. Accordingly, the spectral grid becomes increasingly shifted in $k_x$, such that eventually waves with $k_x\sim 0$ cannot even be resolved. To avoid this issue, we periodically remap the solutions to a regular grid in $(x,y)$ using
\begin{gather}
w = \mathcal{F}_y^{-1}[\mathcal{F}_Y(w_0)e^{-ik_yxL_y/L_x}],
\end{gather}
where $\mathcal{F}_Y$ denotes discrete Fourier transform in $Y$ and $\mathcal{F}_y^{-1}$ denotes inverse discrete Fourier transform in $y$. The remapping needs to be performed whenever $L_xSt \mod L_y =0$, that is, when the system happens to be periodic also in $(x,y)$. This approach can then be considered as semi-Lagrangian due to the periodic remapping. More detailed discussions on such semi-Lagrangian spectral methods to shearing-box simulations can be found in Ref.\,\cite{Barranco2006}.

%

\end{document}